\documentclass[12pt,oneside]{article}
\usepackage{amssymb,amsmath,latexsym,amsthm,amsfonts}
\usepackage[english]{babel}
\usepackage{color,hyperref}
\usepackage{slashbox}
\usepackage{multirow}
\usepackage{graphics}
\usepackage{graphicx}
\newtheorem{prop}{Proposition}

\newtheorem{lem}{Lemma}

\newtheorem{rem}{Remark}

\numberwithin{ex}{section} \numberwithin{rem}{section}
\numberwithin{equation}{section} \numberwithin{thm}{section}
\numberwithin{lem}{section} \numberwithin{coro}{section}

\def\1g{1\hskip -3pt \mbox{l}}

\small\normalsize

\title{Testing instantaneous causality in presence of non constant unconditional variance}

\author{
{\sc Quentin Giai Gianetto
\footnote{IRMAR-INSA, 20 avenue des buttes de Coësmes, CS 70839,
F-35708 Rennes Cedex 7, France. E-mail: Quentin.Giai-Gianetto@insa-rennes.fr }
and Hamdi Ra\"{i}ssi
\footnote{IRMAR-INSA, 20 avenue des buttes de Coësmes, CS 70839,
F-35708 Rennes Cedex 7, France. E-mail: hamdi.raissi@insa-rennes.fr }}}

\begin{document}

\maketitle  \noindent {\em Abstract:} The problem of testing
instantaneous causality between variables with time-varying
unconditional variance is investigated. It is shown that the classical tests
based on the assumption of stationary processes must be avoided in
our non standard framework. More precisely we underline that the
standard test does not control the type I errors, while the tests
with White (1980) and Heteroscedastic Autocorrelation Consistent
(HAC) corrections can suffer from a severe loss of power when the
variance is not constant. Consequently
a modified test based on a bootstrap procedure is proposed. The relevance of the modified test is underlined through a simulation study. The tests considered in this paper are also compared by investigating the instantaneous causality relations between US macroeconomic variables.

\vspace*{.2cm} \noindent {\em Keywords:} VAR model; Unconditionally heteroscedastic errors;
Instantaneous causality.\\

\section {Introduction}
\label{S1}

The concept of causality defined by Granger (1969) is widely used
to analyze cause and effect relationships between macroeconomic and financial variables (see \textit{e.g.} Sims (1972),
Ashenfelter and Card (1982), Hamilton (1983), Lee (1992), Hiemstra
and Jones (1994), Renault and Werker (2005), Gelper and Croux (2007)). The Granger causality
has also been studied in others areas : neuroscience (see
\textit{e.g.} Brovelli et al. (2004), Seth (2008)), gene networks
(Fujita et al. (2009)), geophysics (Reichel, Thejll and Lassen
(2001)), or sociology (Deane and Gutmann (2003)) are some
application domains among others. Causality relationships are often
analyzed by taking into account only the past values of studied
variables. In many situations the prediction of the
unobserved current variables $X_{2t}$ can however be improved by including the
available current information of variables $X_{1t}$. In such a case
the $\textit{instantaneous}$ causality relation between $X_{1t}$ and $X_{2t}$
is investigated (see L\"{u}tkepohl (2005, p 42)).

In the stationary VAR processes framework, the instantaneous
causality is usually tested by using Wald tests for zero restrictions
on the innovation's covariance matrix. Standard tools available in
the commonly used softwares (see L\"{u}tkepohl and Kr\"{a}tzig
(2004)) are based on the assumption of i.i.d. Gaussian innovations.
The weight matrix of the test statistic has to be corrected by using
the White type covariance matrix when the error process is
assumed i.i.d. but non Gaussian (see White (1980)).
In some cases models which produce nonlinear stationary processes
are considered for the error terms as the GARCH or All-Pass models
(see \textit{e.g.} Bauwens, Laurent and Rombouts (2006) or Andrews, Davis and
Breidt (2006)). These models allow to take into account some
dependence in the innovations but also suppose that the
unconditional variance of the innovations process is constant. In order to get a
standard asymptotic distribution of the Wald test statistic in these
situations, Heteroscedasticity and Autocorrelation Consistent (HAC)
corrections can be used (see Den Haan and Lievin (1997) for the HAC
estimation).

Nevertheless many applied papers questioned the assumption of a
constant unconditional variance structure. For instance
Sensier and van Dijk (2004) found that most of the 214 U.S.
macroeconomic variables they investigated exhibit a break in their
unconditional variance. Ramey and Vine (2006) highlighted a
declining variance of the U.S. automobile production and sales. Mc-Connell
and Perez-Quiros (2000) documented a break in variance in the U.S. GDP
growth and pointed out that neglecting non constant variance can
be misleading for the data analysis. It emerges from these studies
that processes with non constant unconditional variance are common features in
practice. All these observations led us to consider the case of
instantaneous causality relationships where the unconditional
variance of the structural innovations changes over time.

Numerous tools for time series analysis in presence of non constant
variance have been proposed in the literature. For instance Tsay
(1988), Horv\'{a}th, Kokoszka and Zhang (2006) or Sanso, Arago and
Carrion (2004) proposed tests for detecting unconditional variance changes in
several situations. Kokoszka and Leipus (2000) and Dahlhaus and Rao
(2006) studied ARCH processes with non constant unconditional
variance. Robinson (1987), Hansen (1995), Francq and Gautier (2004)
or Xu and Phillips (2008) among other references investigated
univariate linear models allowing for a non constant variance.
St\u{a}ric\u{a} (2003) considered a deterministic non constant
specification for the unconditional variance of stock returns, and
noted that such an approach can perform as well as the stationary
GARCH(1,1) model. Kim and Park (2010) studied cointegrated systems
with non constant variance. Bai (2000), Qu and Perron (2007), and
Patilea and Ra\"{\i}ssi (2012) among others investigated the
estimation of multivariate models with time-varying variance. Aue,
H\"{o}rmann, Horv\`{a}th and Reimherr (2009) proposed a test
procedure for detecting variance breaks in multivariate time
series.

In this paper we focus on the test of zero restriction on the
time-varying variance structure. We highlight that the standard Wald
test for instantaneous causality implemented in the commonly used
softwares does not provide suitable critical values when the
variance structure is time-varying. It is also
established that the tests based on White or HAC corrections of the
Wald test statistic can suffer from a severe loss power in certain
important situations. More precisely these tests may be unable to
detect some important alternatives as periodic changes or when the
covariance structure is close to zero, so that its sign likely changes. Noting that the previous tests are not intended to
handle data with non constant unconditional variance, a new approach
for testing instantaneous causality taking into account
non-stationary unconditional variance is proposed in this paper. It is however found that the
asymptotic distribution of the modified statistic is non standard
involving the unknown non constant variance structure in a
functional form. When the asymptotic distribution is non standard,
the wild bootstrap method is widely used in the literature for the
analysis of time series possibly displaying (unconditional)
heteroscedasticity/dependence (see \textit{e.g.}
Gon\c{c}alvez and Kilian (2004), Horowitz, Lobato, Nankervis and
Savin (2006) or Inoue and Kilian (2002)). Therefore a wild bootstrap procedure is provided for
testing zero restrictions on the non constant variance structure. It
is established through theoretical and empirical results that the
modified test is preferable to the tests based on the spurious
assumption of constant unconditional variance.

The plan of the paper is as follows. In the next section, we
introduce the VAR models with non constant variance. In section
\ref{S3} the testing problem for instantaneous causality between
subvectors of a VAR process with non constant variance is discussed.
The asymptotic properties of the tests based on the assumption of
constant unconditional variance are presented. It emerges from this
part that this kind of tests should be avoided in our non standard
framework. As a consequence a test based on the wild bootstrap
procedure taking into consideration non constant variance is
built. The finite sample properties of the tests are investigated
in Section \ref{S4} by Monte Carlo experiments. We also consider US macroeconomic data to illustrate our findings. In section \ref{conclusion},
we draw up a conclusion on our results.

\section {Vector autoregressive model with non constant variance}
\label{S2}

Consider the following VAR model
\begin{eqnarray}\label{VAR}
&&{X}_t={A}_{01}{X}_{t-1}+\dots+{A}_{0p}{X}_{t-p}+u_t\\&&
u_t=H_t\epsilon_t,\nonumber
\end{eqnarray}
where $X_{t}\in\mathbb{R}^{d}$ and it is assumed that
$X_{-p+1},\dots,X_0,X_1,\dots,X_T$ are observed. The $d\times d$
dimensional matrices $A_{0i}$ are such that $\det {A}(z)\neq 0$ for
all $|z|\! \leq \! 1$, where ${A}(z) \!=\!
I_d-\sum_{i=1}^{p}\!{A}_{0i} z^i$ with $I_d$ the $d\times d$ identity matrix. 
Note that the
process $(X_t)$ should be formally written in a triangular form, but
the double subscript is suppressed for notational simplicity. In the following assumption we give the structure of the variance
by using the rescaling approach of Dahlhaus (1997). $\mathcal{F}_{t}$ corresponds to the $\sigma$-field generated by
$\{\epsilon_k:k\leq t\}$ and $\parallel
.\parallel_r$ is such that $\parallel
x\parallel_r:=(E\parallel x\parallel^r)^{1/r}$ for a random variable
$x$ with $\parallel .\parallel$ the Euclidean norm.\\

\textbf{Assumption A1:} \quad (i) The $d\times d$ matrices $H_{t}$
are lower triangular nonsingular matrices with positive diagonal
elements and satisfy $H_{t}=G(t/T)$, where the components of the
matrix $G(r):=\{g_{kl}(r)\}$ are measurable deter\-ministic
functions on the interval $(0,1]$, such that
$\sup_{r\in(0,1]}|g_{kl}(r)|<\infty$, and each $g_{kl}$ satisfies a
Lipschitz condition piecewise on a finite number of sub-intervals partitioning $(0,1]$.
The matrices $\Sigma(r)=G(r)G(r)'$ are assumed positive definite for all $r$ in $(0,1]$.\\
(ii) The process $(\epsilon_t)$ is $\alpha$-mixing and such that
$E(\epsilon_t\mid \mathcal{F}_{t-1})=0$,
$E(\epsilon_t\epsilon_t'\mid \mathcal{F}_{t-1})=I_d$ and
$\sup_t\parallel\epsilon_{t}\parallel_{4\mu}<\infty$ for some
$\mu>1$.\\

If we suppose the process $(\epsilon_t)$ Gaussian and that the
functions $g_{kl}(.)$ are constant we retrieve the standard case.
Nevertheless when the unconditional
variance is time-varying, it can be expected that the
tools developed in the stationary framework are not valid or can suffer from drawbacks since the tests for instantaneous causality are directly
based on the variance structure.
From the piecewise Lipschitz condition
abrupt breaks as well as smooth changes are allowed
for the unconditional variance.
In particular the variance may have a periodic behaviour. The
framework given by our assumption is similar to that of numerous
papers in the literature and encompasses the case of piecewise constant
variance structure (see Pesaran and Timmerman (2004), Hamori and
Tokihisa (1997) or Xu and Phillips (2008)
and references therein). However since we assumed that
$E(\epsilon_t\epsilon_t'\mid \mathcal{F}_{t-1})=I_d$, the error
terms cannot display GARCH effects (as for instance second order
correlation). Therefore the tools proposed in this paper have to be
preferably used for relatively low frequency variables
for which it is commonly admitted that there is no second order
dynamics (for instance monthly, quarterly or annual macroeconomic data, see
Section \ref{realdata} below). In such a situation adding a multivariate GARCH
structure to our model as in Hafner and Linton (2010) can be viewed as too elaborated. The tests proposed in Patilea and Ra\"{\i}ssi (2012) can be used to check
if there is no second order dynamics within the data.
In the framework of {\bf A1} we are interested in testing zero restrictions on the variance structure $\Sigma(r)$.


Now re-write model (\ref{VAR}) as follows

\begin{eqnarray*}
&&{X}_t=(\tilde{X}_{t-1}'\otimes I_d)\theta_0+u_t\\&&
u_t=H_t\epsilon_t,\nonumber
\end{eqnarray*}
with $\tilde{X}_{t-1}=(X_{t-1}',\dots,X_{t-p}')'$ and
$\theta_0=\mbox{vec}\{(A_{01},\dots,A_{0p})\}$ where vec(.) is the
usual column vectorization operator of a matrix and $\otimes$ is the usual Kronecker product. The parameter
vector $\theta_0$ may be estimated by Ordinary Least Squares (OLS)
or Adaptive Least Squares (ALS). Properties of these estimators
are established in Patilea and Ra\"{\i}ssi (2012) under {\bf A1}.
Denoting by $\hat{\theta}$ the ALS (or alternatively OLS) estimator
of $\theta_0$, we introduce the residuals
$\hat{u}_t=X_t-(\tilde{X}_{t-1}'\otimes I_d)\hat{\theta}$. Note that
the more efficient ALS estimation method should be preferred for
approximating the innovations. Otherwise in practice the
autoregressive order is unknown, but it is important to ensure that
the lag length is well adjusted for the analysis of the
instantaneous causality. For instance if the lag length is chosen
too small we get correlated residuals with distorted covariance
structure. On the other hand a too large autoregressive order would imply a large number of parameters to estimate in our
multivariate model. The goodness-of-fit of model
(\ref{VAR}) can be checked by using portmanteau tests proposed in Patilea and
Ra\"{\i}ssi (2011) in our non standard framework. Hence the lag
length is assumed well fitted in the sequel. More particularly the OLS and ALS estimators are $\sqrt{T}$-asymptotically normal if the lag length is well adjusted.

Let $X_{1t}$ and $X_{2t}$ be the subvectors of $X_t:=(X_{1t}',X_{2t}')'$ with respective dimensions $d_1$ and $d_2$
and let $\Sigma_{t}^{12}$ be the $d_1\times d_2$-dimensional
upper right block of $\Sigma_{t}:=E(u_{t}u_{t}')$. Our goal is to determine if it exists an instantaneous causality relation between $X_{1t}$ and $X_{2t}$. The next lemma gives some preliminary results and requires to introduce some additional notations. Let
$\hat{u}_t:=(\hat{u}_{1t}',\hat{u}_{2t}')'$, $\hat{\vartheta}_t:=\hat{u}_{2t}\otimes \hat{u}_{1t}$,
$\hat{v}_t:=\mbox{vec}(\hat{u}_{1t}\hat{u}_{2t}'-\Sigma_{t}^{12})
=\hat{u}_{2t}\otimes \hat{u}_{1t}-\mbox{vec}(\Sigma_{t}^{12})$,
$H_t:=(H_{1t}',H_{2t}')'$ and $G(r):=(G_{1}(r)',G_{2}(r)')'$ be in line
with the partition of $X_t$. We denote by
$[z]$ the integer part of a real number $z$. We also denote by $\Rightarrow$ the
convergence in distribution and $\rightarrow$ the convergence in
probability.

\begin{lem}\label{lem}
Under {\bf A1} we have as $T\to\infty$
\begin{equation}\label{res0}
T^{-1}\sum_{t=1}^T\hat{\vartheta}_t\rightarrow\int_0^1\mbox{vec}(\Sigma^{12}(r))dr,
\end{equation}
and
\begin{equation}\label{res1}
T^{-\frac{1}{2}}\sum_{t=1}^T\hat{v}_t\Rightarrow\mathcal{N}(0,\Omega),
\end{equation}
where
$$\Omega=\int_0^1\left(G_2(r)\otimes G_1(r)\right)\mathcal{M}\left(G_2(r)\otimes G_1(r)\right)'dr
-\int_0^1\mbox{vec}(\Sigma^{12}(r))\mbox{vec}(\Sigma^{12}(r))'dr$$
and
$\mathcal{M}=E(\epsilon_t\epsilon_t'\otimes\epsilon_t\epsilon_t')$.
In addition we also have
\begin{equation}\label{res2}
T^{-\frac{1}{2}}\sum_{t=1}^{[Ts]}\hat{v}_t\Rightarrow\int_0^s(G_2(r)\otimes G_1(r))dB_{\tilde{\Omega}}(r)
\end{equation}
where $B_{\tilde{\Omega}}(.)$ is a Brownian Motion with covariance
matrix
$\tilde{\Omega}:=E(\epsilon_t\epsilon_t'\otimes\epsilon_t\epsilon_t')
-\mbox{vec}(I_d)\mbox{vec}(I_d)',$ with $s\in[0,1]$.
\end{lem}

\vspace*{.7cm}
Using (\ref{res1}) and (\ref{res2}) we shall discuss the test for instantaneous causality
between $X_{1t}$ and $X_{2t}$ assuming spuriously that the unconditional variance is constant and propose a new test adapted to our
framework in the next section. Nevertheless remarks on the result (\ref{res1}) must be made.

\begin{rem}\label{remarque1}
 Let be

$$\Sigma(r)=\left(
              \begin{array}{cc}
                \Sigma^{11}(r) & \Sigma^{12}(r) \\
                \Sigma^{21}(r) & \Sigma^{22}(r) \\
              \end{array}
            \right)
$$
in line with the partition of $X_t$. If we suppose that
$\Sigma^{12}(r)=0$ for all $r\in(0,1]$, which corresponds to the
case of no instantaneous causality relation between $X_{1t}$
and $X_{2t}$ (see the null hypothesis $H_0$ below), it follows that
$\hat{v}_t=\hat{\vartheta}_t=\hat{u}_{2t}\otimes\hat{u}_{1t}$ and

$$\Omega=\int_0^1\left(G_2(r)\otimes G_1(r)\right)\mathcal{M}\left(G_2(r)\otimes G_1(r)\right)'dr.$$
In such a situation from Lemmas \ref{lem1}, \ref{precedent},
\ref{lemtcllln} and the proof of Lemma \ref{lem}, it is clear that

\begin{equation}\label{omegaw}
\hat{\Omega}_w:=T^{-1}\sum_{t=1}^T\hat{u}_{2t}\hat{u}_{2t}'\otimes
\hat{u}_{1t}\hat{u}_{1t}'\rightarrow\Omega.
\end{equation}
In particular when the process
$(u_t)$ is assumed Gaussian with non constant variance
the expression of $\Omega$ simplifies itself
into $\int_0^1\Sigma^{22}(r)\otimes\Sigma^{11}(r)dr$ by using (\ref{remark}).
\end{rem}
\begin{rem} Assume that the process $(u_t)$ is i.i.d. Gaussian with
(constant) variance noted
$$\Sigma_{u}=\left(
\begin{array}{cc}
\Sigma_{u}^{11} & 0 \\
0 & \Sigma_{u}^{22} \\
\end{array}
\right),
$$ which corresponds to the standard case under the hypothesis $H_0$
below. In such a case the expression of $\Omega$ simplifies itself into
$\Sigma_{u}^{22}\otimes\Sigma_{u}^{11}$. Under the strong assumption
of i.i.d. Gaussian error process it can be shown that
\begin{equation}\label{omegas}
\hat{\Omega}_{st}:=\left\{\left(T^{-1}\sum_{t=1}^T
\hat{u}_{2t}\hat{u}_{2t}'\right)\otimes
\left(T^{-1}\sum_{t=1}^T\hat{u}_{1t}\hat{u}_{1t}'\right)\right\}
\rightarrow\Sigma_{u}^{22}\otimes\Sigma_{u}^{11}.
\end{equation}

\end{rem}

\vspace*{.7cm}

\section{Testing for instantaneous causality}
\label{S3}

In the sequel we follow the notations of L\"{u}tkepohl (2005). Denote
by $X_{2t}(1|\{X_k|k < t\})$ the optimal one step linear predictor
of $X_{2t}$ at the date $t-1$, based on the information of the past
of the process $(X_t)$.
Similarly we define the one step linear predictor $X_{2t}(1|\{X_k|k
< t\}\cup\{X_{1t}\})$ based on the past of $(X_t)$ and the present
of $(X_{1t})$.
It is said that there is no instantaneous linear causality between $(X_{1t})$ and $(X_{2t})$ if

$$X_{2t}(1|\{X_k|k < t\}\cup\{X_{1t}\})=X_{2t}(1|\{X_k|k < t\}).$$
In the case of non constant variance following the assumption {\bf A1} and, more particularly, because we assumed that the $H_{t}$'s are lower triangular
nonsingular matrices with positive diagonal elements, it can be
shown that there is no instantaneous causality between $X_{1t}$ and
$X_{2t}$ if and only if the $\Sigma_t^{12}$'s are all equal to zero
following the same arguments to those in L\"{u}tkepohl (2005, pp
46-47). Consequently in our non standard framework the following pair
of hypotheses has to be tested:

$$H_0:\:\Sigma^{12}(r)=0\:\mbox{vs}\:H_1:\:\Sigma^{12}(r)\neq0
\:\mbox{for}\: r\in[a,b]\subseteq[0,1]\: \mbox{with fixed}\: a<b.$$
Now if we consider
the case where the variance is assumed constant
$\Sigma_t=\Sigma_{u}$ for all $t$, it is well known that there is no
instantaneous causality between $X_{2t}$ and $X_{1t}$ if and only if
$\Sigma_{u}^{12}=0$ with obvious notation. Therefore the following
pair of hypotheses is tested under standard assumptions:

$$H_0':\:\Sigma_{u}^{12}=0\quad\mbox{vs}\quad H_1':\:\Sigma_{u}^{12}\neq0,$$
The block $\Sigma_{u}^{12}$ is usually estimated by
$T^{-1}\sum_{t=1}^T\hat{u}_{1t}\hat{u}_{2t}'$ which converges in
probability to $\int_0^1\Sigma^{12}(r)dr$ under {\bf A1}. Hence such hypothesis testing does not take into account the
time-varying variance in the sense that it can only be interpreted
as a \textit{global} zero restriction testing of the covariance
structure, i.e. testing $\int_0^1\Sigma^{12}(r)dr=0$ against the
alternative
$\int_0^1\Sigma^{12}(r)dr\neq0$. 
Then $H_0'$ and $H_1'$ are inappropriate for testing instantaneous
causality in our non standard framework.

It is interesting to point out that $H_0$ is a particular case of
$H_0'$, i.e. $H_0\subset H_0'$, since $H_0'$ corresponds to
$\int_0^1\Sigma^{12}(r)dr=0$. On the other hand since
$\int_0^1\Sigma^{12}(r)dr\neq0$ implies that $\Sigma^{12}(r)\neq0$,
then $H_1'\subset H_1$. More precisely, if $\Sigma^{12}(r)\neq0$ for $r\in[a,b]\subseteq[0,1]$, we may have
either $\int_0^1\Sigma^{12}(r)dr\neq0$, which corresponds to $H_1\cap H_1'$, or $\int_0^1\Sigma^{12}(r)dr=0$, which
corresponds to $H_1\cap H_0'$. Note that we have $H_1=(H_1\cap
H_1')\cup(H_1\cap H_0')$ and $(H_1\cap H_1')\cap(H_1\cap
H_0')=\varnothing$. It is shown in the next part that the case
$H_1\cap H_0'$ entails a loss of power for tests built on
the assumption of constant unconditional variance of the innovations.\\


\subsection{Tests based on the assumption of constant error variance}

In this section the consequences of non constant variance on the instantaneous causality tests based on the spurious assumption of
a stationary process are analyzed. Let be
$\delta_T:=T^{-\frac{1}{2}}\sum_{t=1}^T\hat{\vartheta}_t$ where we
recall that $\hat{\vartheta}_t=\hat{u}_{2t}\otimes \hat{u}_{1t}$.
The standard test statistic is given by

\begin{equation*}
S_{st}=\delta_T'(\hat{\Omega}_{st})^{-1}\delta_T,
\end{equation*}
where $\hat{\Omega}_{st}$ is defined in (\ref{omegas}). Under {\bf A1}
it can be shown that $\hat{\Omega}_{st}\rightarrow\int_0^1
\Sigma^{22}(r)dr\otimes\int_0^1\Sigma^{11}(r)dr=:\Omega_{st}$ and we
obviously have $\Omega\neq\Omega_{st}$ in general.

If the practitioner (spuriously) assumes that the error process is
iid but not Gaussian, $u_{1t}$ and $u_{2t}$ could be dependent and
the following statistic with White type correction should be used:

\begin{equation*}
S_w=\delta_T'(\hat{\Omega}_w)^{-1}\delta_T,
\end{equation*}
where the weight matrix $\hat{\Omega}_w$ is defined in
(\ref{omegaw}). Recall that $\hat{\Omega}_w$ is a consistent
estimator of $\Omega$ under $H_0$ and
it is clear from the proof of Lemma \ref{lem} that 

\begin{equation}\label{statwhite}
\hat{\Omega}_w\rightarrow\Omega+
\int_0^1\mbox{vec}(\Sigma^{12}(r))\mbox{vec}(\Sigma^{12}(r))'dr
\end{equation}
under $H_1$.

Finally the practitioner may again (spuriously) suppose that the
error process is stationary and that the observed heteroscedasticity
is a consequence of the presence of nonlinearities. However note
that the assumed heteroscedasticity is only conditional while the
unconditional variance is still constant in this case. This kind of
situation can arise if we (spuriously) assume that the innovations
process is driven by a GARCH model or any other model displaying
nonlinearities such as models driven by hidden Markov chains or
All-Pass models (see Amendola and Francq (2009)). Note that the test
proposed in Sanso, Arago and Carrion (2004) could be used to detect
changes in the unconditional variance. In such a case HAC type
weight matrices should be used in the test statistic. For simplicity
we focus on the VARHAC weight matrix (see Den Haan and Levin
(1997)).
Denote by
$\mathcal{\hat{A}}_{m,1},\dots,\mathcal{\hat{A}}_{m,m}$ the
coefficients of the LS regression of $\hat{\vartheta}_t$ on
$\hat{\vartheta}_{t-1},\dots,\hat{\vartheta}_{t-m}$, taking
$\hat{\vartheta}_{t}=0$ for $t\leq0$. Introduce $\hat{z}_{m,t}$ the
residuals of such a regression and
$\hat{\Omega}_h=\mathcal{A}(1)^{-1}\hat{\Sigma}_z\mathcal{A}(1)^{-1}$
where
$\mathcal{A}(1)=I_{d_1d_2}-\sum_{k=1}^m\mathcal{\hat{A}}_{m,k}$ and
$\hat{\Sigma}_z=T^{-1}\sum_{t=1}^T\hat{z}_{m,t}\hat{z}_{m,t}'$. The
order $m$ can be chosen by using an information criterion. The
following statistic involving VARHAC type weight matrix may be used

\begin{equation*}
S_h=\delta_T'(\hat{\Omega}_h)^{-1}\delta_T.
\end{equation*}
Since we assumed that the autoregressive
order $p$ is well adjusted (or known), the process
$\vartheta_t=u_{2t}\otimes u_{1t}$ is uncorrelated and it can be
shown that the $\mathcal{A}_{m,k}$'s converge to zero in
probability. Therefore $\hat{\Omega}_h\rightarrow\Omega$ under $H_0$
and

\begin{equation}\label{stathac}
\hat{\Omega}_h\rightarrow\Omega+
\int_0^1\mbox{vec}(\Sigma^{12}(r))\mbox{vec}(\Sigma^{12}(r))'dr,
\end{equation}
under $H_1$, so that $\hat{\Omega}_h$ and $\hat{\Omega}_w$ are asymptotically equivalent
in the framework of {\bf A1}. This is not surprising since second order dynamics are in fact excluded
in {\bf A1}.\\ 

In this part the asymptotic properties of the above statistics are investigated. The asymptotic behavior of the statistics in our non standard
framework is first established under $H_0$. The results are direct
consequences of (\ref{res1})

\begin{prop}\label{prop2}
Assume that $H_0$ hold. Then under {\bf A1} we have as $T\to\infty$
\begin{equation}\label{resprop1}
S_{st}\Rightarrow\sum_{j=1}^{d_1d_2}\lambda_jZ_j^2,
\end{equation}
where the $Z_j$'s are independent $\mathcal{N}(0,1)$ variables, and
$\lambda_1,\dots,\lambda_{d_1d_2}$ are the eigenvalues of the matrix
$\Omega_{st}^{-\frac{1}{2}}\Omega\Omega_{st}^{-\frac{1}{2}}$.
In addition we also have
\begin{equation}\label{resprop2}
S_w\Rightarrow\chi_{d_1d_2}^2\quad\mbox{and}\quad S_h\Rightarrow\chi_{d_1d_2}^2.
\end{equation}
\end{prop}

For a fixed asymptotic level $\alpha$, the standard test ($W_{st}$
hereafter) consists in rejecting the hypothesis of no instantaneous
causality between $X_{1t}$ and $X_{2t}$ if $S_{st}>\chi^2_{d_1d_2,
1-\alpha}$ where $\chi^2_{d_1d_2, 1-\alpha}$ is the $(1-\alpha)th$
quantile of the $\chi^2_{d_1d_2}$ law. Therefore it appears from
(\ref{resprop1}) that the standard test is not able to control the
type I error since $\Omega_{st}\neq\Omega$ in general. Denote by $W_w$
(resp. $W_h$) the test consisting to reject the hypothesis of
no instantaneous causality  if $S_w>\chi^2_{d_1d_2, 1-\alpha}$
(resp. $S_h>\chi^2_{d_1d_2, 1-\alpha}$). From (\ref{resprop2}) we
see that the $W_w$ and $W_h$ tests should have good size properties
for large enough $T$. 

Now we turn to the study of the power properties of the $W_{st}$, $W_w$ and $W_h$
tests in the cases $H_1\cap H_1'$ and $H_1\cap H_0'$. We
first consider the situation $H_1\cap H_1'$, which corresponds to
the case $\int_0^1\Sigma^{12}(r)dr\neq0$. From (\ref{res0}) it is
clear that $T^{-1}S_i$, with $i=st,w,h$, converge in probability to
strictly positive constants if $\int_0^1\Sigma^{12}(r)dr\neq0$. It
follows that the $S_{st}$, $S_w$ and $S_h$ statistics grow to infinity
as fast as $T$. Therefore we can expect that the tests based on the
assumption of constant variance will detect a possible instantaneous
causality for large enough sample sizes when $H_1\cap H_1'$ hold.
The abilities of the $W_{st}$, $W_w$ and $W_h$ tests to detect the case
$\int_0^1\Sigma^{12}(r)dr\neq0$ are compared considering the
approximate Bahadur slope approach (Bahadur (1960)). 
For the test
based on the $S_{st}$ statistic define
$q_{st}(x)=-\log P_0\left(S_{st}>x\right)$ for any
$x>0$, where $P_0$ stands for the limit distribution
of $S_{st}$ under $\Sigma^{12}(r)=0$.

For a fixed alternative such that
$\varpi=\int_0^1\Sigma^{12}(r)dr\neq0$, consider the asymptotic
slope $c_{st}(\varpi) = 2\lim_{T\rightarrow\infty} T^{-1} q_{st}(S_{st})$.
Define similarly $c_w(\varpi)$ and $c_h(\varpi)$ for the $W_w$,
$W_h$ tests and also the asymptotic relative efficiencies
$ARE_{S_w,S_{st}}(\varpi) = c_w(\varpi)/c_{st}(\varpi)$ and
$ARE_{S_h,S_{st}}(\varpi) = c_h(\varpi)/c_{st}(\varpi)$. A relative
efficiency $ARE_{S_h,S_{st}}(\varpi)\geq 1$ suggests that the $W_h$
test is more able to detect the case $\int_0^1\Sigma^{12}(r)dr\neq0$
than the test based on the $S_{st}$ statistic. In such a case the
$p$-values of
the $W_h$ test converge faster to zero than those of the $W_{st}$ test.\\

\begin{prop} \label{bahadur}
Under {\bf A1} we have $ARE_{S_w,S_{st}}(\varpi)\geq1$ and
$ARE_{S_h,S_{st}}(\varpi)\geq1$ for every alternative such that
$\varpi=\int_0^1\Sigma^{12}(r)dr\neq0$.\\
\end{prop}

The proof of Proposition \ref{bahadur} is similar to the proof of
Proposition 5.3 in Patilea and Ra\"{\i}ssi (2012) and is then omitted.
When the errors are assumed iid Gaussian we obtain
$ARE_{S_w,S_{st}}(\omega)=1$ and $ARE_{S_h,S_{st}}(\omega)=1$.
Nevertheless if the variance structure of the errors is non constant
with $\int_0^1\Sigma^{12}(r)dr\neq0$, the $W_h$ or $W_w$ tests
achieve a gain in power when compared to the $W_{st}$ test.

Finally we study the ability of the $W_{st}$, $W_w$ and $W_h$ tests in
detecting instantaneous causality when $H_1\cap H_0'$ hold, that is
$\Sigma^{12}(r)\neq0$ and $\int_0^1\Sigma^{12}(r)dr=0$. In this case
$\hat{\Omega}_h=O_p(1)$ and $\hat{\Omega}_w=O_p(1)$ from
(\ref{statwhite}) and (\ref{stathac}) and we also have
$\hat{\Omega}_{st}=O_p(1)$, while the non centrality term is
$T^{-\frac{1}{2}}\sum_{t=1}^T\Sigma_t^{12}=o(T^{\frac{1}{2}})$.
Therefore we have $S_i=o_p(T)$ with $i=st,w,h$ in the case $H_1\cap
H_0'$. When such eventuality is considered, it is clear that the
tests based on the assumption of stationary errors may suffer from a
severe loss of power. This is a consequence of the fact that this
kind of tests are not intended to take into account
time varying variance. The case $H_1\cap H_0'$ can arise in the
important case where $\Sigma^{12}(r)\neq0$ but close to zero so that
$\Sigma^{12}(r)$ may have a changing sign. Even when at least one of
the components of $\Sigma^{12}(r)$ is far from zero, we can have
$\int_0^1\Sigma^{12}(r)dr=0$ as for instance in some cases where the
variance structure is periodic. This can be seen by considering the
bivariate case and taking $\Sigma^{12}(r)=c\cos(\pi r)$ or
$\Sigma^{12}(r)=c\mathbf{1}_{[0,\frac{1}{2}]}(r)-c\mathbf{1}_{]\frac{1}{2},1]}(r)$
with $c\in\mathbb{R}$. Therefore the tests based on the spurious
assumption of constant unconditional variance for the error must be
avoided.\\ 

In summary it is found that the tests based on the White and VARHAC
corrections should control well the type I errors for large enough samples on
the contrary to the $W_{st}$. In addition it appears that in the case
of non constant unconditional variance and when $H_1\cap H_1'$ hold,
the $W_h$ and $W_w$ tests have better power properties than the
$W_{st}$ test. Therefore the $W_h$ and $W_w$ tests should be preferred
to the $W_{st}$ test when the unconditional variance is time-varying.
However it is also found that the tests based on the assumption of
constant variance may suffer from a severe loss of power in the
important cases where $\int_0^1\Sigma^{12}(r)dr=0$ (or
$\int_0^1\Sigma^{12}(r)dr\approx0$). A bootstrap test
circumventing this power problem in the case $H_1\cap H_0'$ is proposed
in the next part.

\subsection{ A bootstrap test taking into account non constant variance}
Introduce $\delta_s=T^{-\frac{1}{2}}\sum_{t=1}^{[Ts]}\hat{\vartheta}_t$
with $s\in[0,1]$
and consider the following statistic:

$$S_b=\sup_{s\in[0,1]}
||\delta_s||_2^2.
$$
Under $H_0$ and from (\ref{res2}) we write:

$$\delta_{s}\Rightarrow\int_0^s(G_2(r)\otimes G_1(r))dB_{\tilde{\Omega}}(r):=K(s),$$
where the covariance matrix becomes
$\tilde{\Omega}=\mathcal{M}=E(\epsilon_t\epsilon_t'\otimes\epsilon_t\epsilon_t')$.
Therefore under $H_0$ we have from the Continuous Mapping Theorem

\begin{equation}\label{statsup}
S_b\Rightarrow\sup_{s\in[0,1]}||K(s)||_2^2,
\end{equation}
since the functional $f(Y)=\sup_{s\in[0,1]}||Y(s)||_2^2$ is
continuous for any $Y\in D[0,1]$, the space of c\`{a}dl\`{a}g
processes on [0,1].
Under $H_1$ we obtain

\begin{equation}\label{zitoun}
T^{-\frac{1}{2}}\delta_{s}=T^{-1}\sum_{t=1}^{[Ts]}\hat{v}_t
+T^{-1}\sum_{t=1}^{[Ts]}\mbox{vec}(\Sigma_t^{12})
\end{equation}
with $\tilde{\Omega}$ defined in (\ref{res2}). The first term in
the right hand side of (\ref{zitoun}) converges to zero in
probability, while we have
$T^{-1}\sum_{t=1}^{[Ts]}\mbox{vec}(\Sigma_t^{12})=\int_0^s\mbox{vec}(\Sigma^{12}(r))dr+o(1)$
and
$\sup_{s\in[0,1]}||\left\{\int_0^s\mbox{vec}(\Sigma^{12}(r))dr\right\}||_2^2=C>0$.
Hence we have in such a situation $S_b=CT+o_p(T)$.



From (\ref{statsup}) we see that the asymptotic distribution of
$S_b$ under the null $H_0$ is non standard and depends on the
unknown variance structure and the fourth order cumulants of the
process $(\epsilon_t)$ in a functional form. Thus the statistic
$S_b$ cannot directly be used to build a test and we consider a wild
bootstrap procedure to provide reliable quantiles for testing the
instantaneous causality.
In the literature such procedures were used for
investigating VAR model specification as in Inoue and Kilian (2002)
among others. The
reader is referred to Davidson and Flachaire (2008) or Gon\c{c}alves
and Kilian (2004, 2007) and references therein for the wild
bootstrap procedure method. For resampling our test statistic we
draw $B$ bootstrap sets given by
$\vartheta_t^{(i)}:=\xi_t^{(i)}\hat{\vartheta}_{t}=\xi_t^{(i)}
\hat{u}_{2t}\otimes\hat{u}_{1t}$,
$t\in\{1,\dots,T\}$ and $i\in\{1,\dots,B\}$, where the univariate
random variables $\xi_t^{(i)}$ are taken iid standard Gaussian, independent from $(u_t)$.
For a given $i\in\{1,\dots,B\}$ set
$\delta_s^{(i)}=T^{-\frac{1}{2}}\sum_{t=1}^{[Ts]}\vartheta_t^{(i)}$ and
$S_b^{(i)}=\sup_{s\in[0,1]}||\delta_s^{(i)}||_2^2$. In our procedure
bootstrap counterparts of the $x_t$'s are not generated and the residuals are directly used to generate the bootstrap
residuals. This is motivated by the fact that zero restrictions are tested on the variance structure,
so that we only consider the residuals in the test statistic. In addition it is seen from (\ref{mvt})
that the residuals and the errors are asymptotically equivalent. It is also clear
that the wild bootstrap method is designed to replicate the pattern
of non constant variance of the residuals in $S_b^{(i)}$. More precisely we have under {\bf A1}

\begin{equation}\label{bootres}
S_b^{(i)}\Rightarrow^P\sup_{s\in[0,1]}||K(s)||_2^2,
\end{equation}
where we denote by $\Rightarrow^P$ the weak convergence in probability. A proof of (\ref{bootres}) is provided in the Appendix. Note
that we have by construction $E^*(\xi_t^{(i)}\hat{\vartheta}_t)=0$ even when the alternative is
true, that is $E(\vartheta_t)\neq0$ (recall that
$\vartheta_t=u_{2t}\otimes u_{1t}$). As a consequence the result (\ref{bootres}) is hold whatever $\Sigma(r)^{12}=0$ or  $\Sigma(r)^{12}\neq0$.

The $W_b$ test consists in rejecting $H_0$ if the statistic $S_b$
exceeds the $(1-\alpha)$ quantile of the bootstrap distribution.
Under $H_1$ with $\int_0^1\Sigma(r)^{12}dr\neq0$ we note that all
the statistics considered in this paper increase at the rate $T$.
However when $\Sigma(r)^{12}\neq0$ with
$\int_0^1\Sigma(r)^{12}dr\approx0$, we can expect that the $W_b$
test is more powerful than the tests based on the assumption of
constant unconditional variance. In such situations we may have
$S_w=o_p(T)$, $S_h=o_p(T)$ while again $S_b=O_p(T)$. If the
unconditional variance is constant, that is
$\Sigma^{12}(r)=\Sigma_u^{12}$, note that $S_w=O_p(T)$, $S_h=O_p(T)$
and $S_b=O_p(T)$. Hence we can expect no major loss of power for the
$W_b$ when compared to the $W_w$ and $W_h$ tests if the underlying
structure of the variance is constant. In general since $S_b=O_p(T)$ and in
view of (\ref{bootres}), the $W_b$ test is consistent.
From the above results we can draw the conclusion that the $W_b$ is
preferable if the unconditional variance is non constant for large
enough sample sizes. 


\section{Numerical illustrations}
\label{S4}

In this section the $W_{b}$ test is compared to the $W_{st}$ and $W_{w}$ tests. The VARHAC statistic being asymptotically equivalent to the White statistic under \textbf{A1}, as noted above, we did not take into account this test in our comparisons. First the type I errors and power properties of the three tests are compared using simulated bivariate VAR(1) processes with unconditional time-varying variance. The tests are next applied to two macroeconomic data sets.

\subsection{Simulation study}

For our experiments we simulated simple bivariate VAR(1) processes where the autoregressive parameters are inspired from those estimated from the money supply and inflation in the U.S. data (see section below). The data generating process can be written as

\begin{equation}\begin{pmatrix}
X_{1,t}\\
X_{2,t}
\end{pmatrix}=\begin{pmatrix}
0.64&-1\\
-0.01&0.44
\end{pmatrix}\begin{pmatrix}
X_{1,t-1}\\
X_{2,t-1}
\end{pmatrix}+\left(
                \begin{array}{c}
                  u_{1,t} \\
                  u_{2,t} \\
                \end{array}
              \right)
\label{eqex}\end{equation}
where the innovations are Gaussian with variance structure $\Sigma(r)$ respecting the assumption $\textbf{A1}$. Two cases are considered for this structure :
\begin{itemize}
\item \textbf{Case 1: Empirical size setting.} It does not exist an instantaneous causality relation between $X_{1,t}$ and $X_{2,t}$ :
\[\Sigma(r)=\begin{pmatrix}
\Sigma^{11}(r) & 0\\
0 & \Sigma^{22}(r)\end{pmatrix}\quad\forall r\in(0,1].\]
where $\Sigma^{11}(r)=a-\cos(br)$ and $\Sigma^{22}(r)=a+\sin(br)$ correspond to the non constant variances of the innovations. We take $a>1$ which represents the level of these variances and $b$ their angular frequency.
\item \textbf{Case 2: Empirical power setting.} It exists an instantaneous causality relation between $X_{1,t}$ and $X_{2,t}$ :
\[\Sigma(r)=\begin{pmatrix}
\Sigma^{11}(r)& \Sigma^{12}(r) \\
\Sigma^{12}(r) & \Sigma^{22}(r)\end{pmatrix}\quad\forall r\in(0,1].\]
where $\Sigma^{12}(r)=c \sin(2\pi r)$ respects the case $\int_0^1\Sigma^{12}(r)dr=0$ with $\Sigma^{12}(r)\neq 0$ almost everywhere on $r\in(0,1]$, and $\Sigma^{11}(.)$, $\Sigma^{22}(.)$ are defined as in Case 1. In particular, the constant $c$ will allow to investigate the ability of our modified test for detecting such alternative when it gets closer to the null hypothesis.
\end{itemize}

Note that $a$, $b$ and $c$ have to be chosen to fulfill the positive definite condition on $\Sigma(r)$ for all $r$ in (0,1]. For instance this property is checked if $a=1.1$, $b=11$ and $\frac{2}{3}\geq c>0$.

\begin{figure}
    \centering
       \includegraphics[scale=0.65]{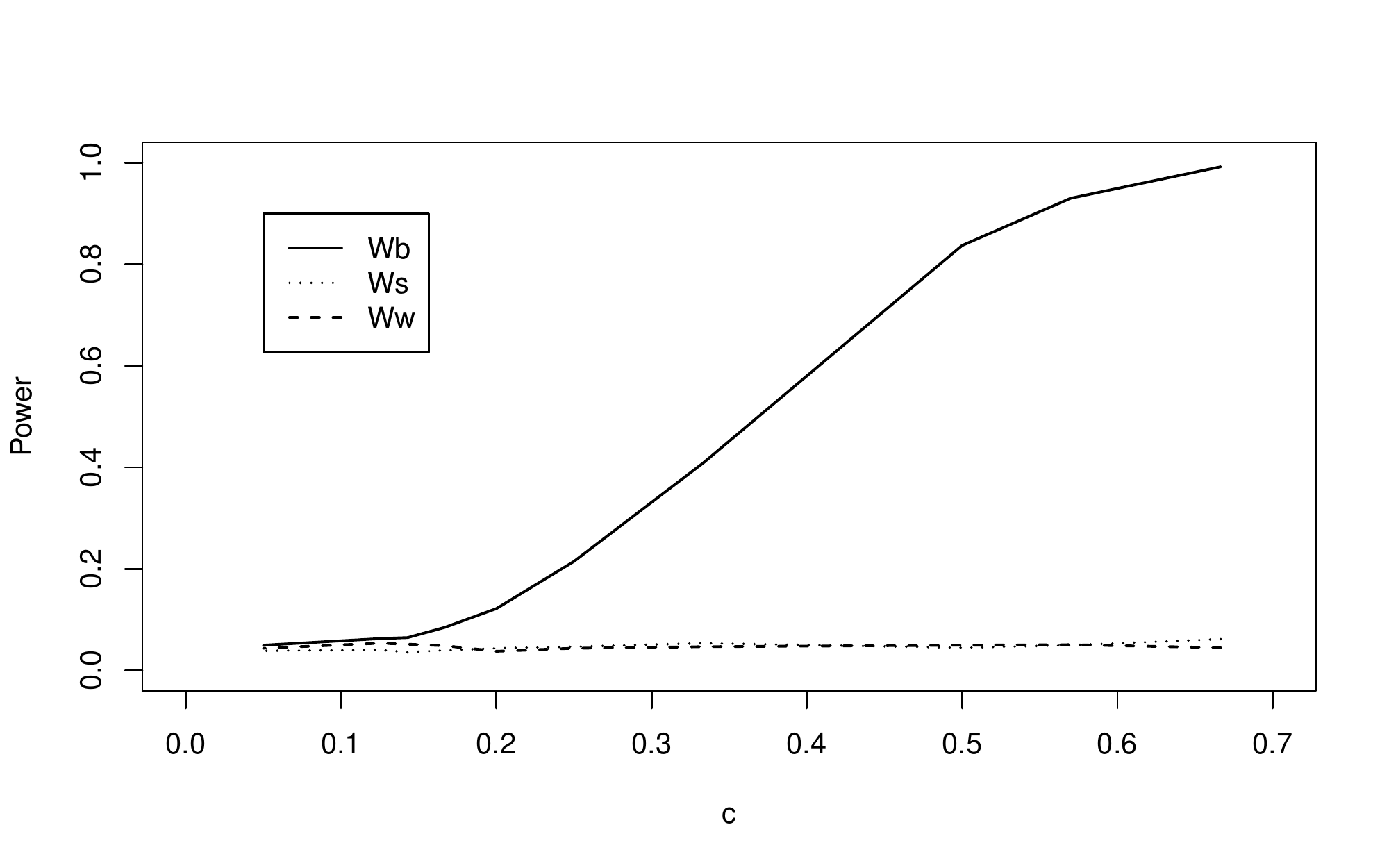}
       \caption{The empirical power of the $W_{b}$, $W_{st}$ and $W_{w}$ tests for fixed sample size $T=500$ and varying $c$ parameter. The asymptotic nominal level is $5\%$ and we take $a=1.1$, $b=11$.}
	\label{EvolPow}
\end{figure}

The finite sample properties of the tests are assessed by means of the following Monte Carlo experiments. For each sample size, 1000 time series following (\ref{eqex}) are generated. The lag length is assumed known and the autoregressive parameters are estimated by using the commonly used OLS method. In all our experiments we use 299 bootstrap iterations for the $W_b$ test. We use processes generated by $\textbf{Case 1}$ to shed light on the control of the type I errors of the studied tests. The results are reported in Table \ref{TabI}. On the other hand processes generated by $\textbf{Case 2}$ are considered for the power study. The results are given in Table
\ref{TabP} and Figure \ref{EvolPow}. Note that in Table \ref{TabI} and Table \ref{TabP} we take $a=1.1$, $b=11$ and $c=0.5$ while in Figure \ref{EvolPow} we take several values for $c$ and $a=1.1$, $b=11$.

\begin{center}
\begin{table}[h!]
\centering
{\small
\begin{tabular}{|c|c|c|c|c|c|c|c|c|c|c|c|c|c|c|}
\cline{3-11}
\multicolumn{2}{c|}{} & \multicolumn{9}{c|}{Asymptotic nominal level}\\
\cline{3-11}
\multicolumn{2}{c|}{} & \multicolumn{3}{c|}{1\%}  & \multicolumn{3}{c|}{5\%} & \multicolumn{3}{c|}{10\%}\\
\cline{3-11}
\multicolumn{2}{c|}{} & $W_{st}$ & $W_{w}$ & $W_{b}$ & $W_{st}$ & $W_{w}$ & $W_{b}$ & $W_{st}$ & $W_{w}$ & $W_{b}$\\
\hline
\multirow{6}{*}{\rotatebox{90}{Sample size}}& 50 & 0.010 & 0.007 & 0.007 & 0.048 & 0.050 & 0.047 & 0.102 & 0.113 & 0.105\\
\cline{2-11}
& 100 & 0.009 & 0.009 & 0.010 & 0.057 & 0.066 & 0.066 & 0.095 & 0.096 & 0.107\\
\cline{2-11}
& 200 & 0.011 & 0.010 & 0.011 & 0.045 & 0.047 & 0.052 & 0.101 & 0.116 & 0.122\\
\cline{2-11}
& 500 & 0.011 & 0.010 & 0.010 & 0.042 & 0.047 & 0.050 & 0.099 & 0.101 & 0.101\\
\cline{2-11}
& 1000 & 0.008 & 0.009 & 0.010 & 0.056 & 0.051 & 0.051 & 0.088 & 0.097 & 0.103\\
\hline
\end{tabular}
}
\caption{The empirical size for the studied tests with asymptotic nominal level
1\%, 5\%, 10\% and $a=1.1$, $b=11$, $c=0.5$.}
\label{TabI}
\end{table}

\begin{table}[h!]
\centering
{\small
\begin{tabular}{|c|c|c|c|c|c|c|c|c|c|c|c|c|c|c|}
\cline{3-11}
\multicolumn{2}{c|}{} & \multicolumn{9}{c|}{Asymptotic nominal level}\\
\cline{3-11}
\multicolumn{2}{c|}{} & \multicolumn{3}{c|}{1\%}  & \multicolumn{3}{c|}{5\%} & \multicolumn{3}{c|}{10\%}\\
\cline{3-11}
\multicolumn{2}{c|}{} & $W_{st}$ & $W_{w}$ & $W_{b}$ & $W_{st}$ & $W_{w}$ & $W_{b}$ & $W_{st}$ & $W_{w}$ & $W_{b}$\\
\hline
\multirow{6}{*}{\rotatebox{90}{Sample size}}& 50 & 0.014 & 0.005 & 0.003 & 0.056 & 0.040 & 0.045 & 0.110 & 0.085 & 0.132\\
\cline{2-11}
& 100 & 0.012 & 0.005 & 0.012 & 0.056 & 0.038 & 0.102 & 0.098 & 0.079 & 0.213\\
\cline{2-11}
& 200 & 0.017 & 0.010 & 0.076 & 0.063 & 0.048 & 0.305 & 0.106 & 0.093 & 0.512\\
\cline{2-11}
& 500 & 0.011 & 0.006 & 0.486 & 0.050 & 0.038 & 0.837 & 0.105 & 0.080 & 0.930\\
\cline{2-11}
& 1000 & 0.015 & 0.011 & 0.966 & 0.056 & 0.045 & 0.997 & 0.108 & 0.088 & 1.000\\
\hline
\end{tabular}
}
\caption{The empirical power for the studied tests based on asymptotic nominal levels
1\%, 5\%, 10\% and $a=1.1$, $b=11$, $c=0.5$.}
\label{TabP}
\end{table}
\end{center}

In our example the $W_{st}$, $W_{w}$ and $W_{b}$ tests seem to control the type I errors reasonably well (see Table \ref{TabI}). We can remark that the standard test provides similar results as compared to the other tests. Nevertheless this outcome does not have to be generalized in view of (\ref{resprop1}). In addition recall from Proposition \ref{bahadur} that the $W_{st}$ is less powerful than the $W_w$ and $W_h$ tests.
Now if we turn to the alternative given by $\textbf{Case 2}$, Table \ref{TabP} clearly shows that the $W_{st}$ and $W_{w}$ tests have no power as the sample sizes increase on the contrary of the $W_{b}$ test. This confirms the theoretical results obtained when $\int_0^1\Sigma^{12}(r)dr\approx 0$. For instance the $W_{b}$ test is almost always rejecting the null hypothesis $H_0$ for a sample size of 1000, while the $W_{st}$ and $W_{w}$ tests are completely not able to detect the alternative in this case.

In the above power experiments the changes of $\Sigma^{12}(r)$ around zero were
fixed by a constant $c$. In this part we illustrate the ability of the tests to
detect departures from the null hypothesis $\Sigma^{12}(r)=0$, while we again have
$\int_0^1\Sigma^{12}(r)dr=0$ in all situations.
Figure \ref{EvolPow} represents the power of the three tests when the parameter
$c$ takes several values, while the sample is fixed $T=500$. We clearly observe that the relative rejection frequencies
of the $W_{b}$ test increases when the covariance structure $\Sigma^{12}(r)\neq 0$
goes away from zero but verifying $\int_0^1\Sigma^{12}(r)dr=0$. On the other hand
we again remark that the relative rejection frequencies of the tests based on the assumption
of constant variance remain close to the asymptotic nominal level
even when $c$ takes large values.

\subsection{Application to macroeconomic data sets}
\label{realdata}
In this part we compare the $W_{st}$ and $W_{w}$ tests with the $W_{b}$ test by investigating instantaneous causality relationships in U.S. macroeconomic data sets.

\subsubsection{Money supply and inflation in the U.S.A.}
The relationship between money supply and inflation is fundamental in the macroeconomic
theories explaining the influence of monetary policy on economy. For instance,
the quantity theory of money assumes a proportional relationship between money
supply and the price level. The reader is referred to Case, Fair and Oster (2011) or
Mankiw and Taylor (2006) concerning the theoretical links which can be made
between money supply and inflation. Many studies investigate this relation
from an empirical point of view. Their results are however ambiguous.
For instance, Turnovsky and Wohar (1984) used a simple macro model to investigate the relationship and find that the rate of inflation is independent of the monetary growth rate in the U.S.A. over the period 1923-1960, while Benderly and Zwick (1985) or
Jones and Uri (1986) give some evidence of relationship over the respective periods 1955-1982 and 1953-1984. Here we investigate the hypothesis of an instantaneous causal
relationship between money supply and inflation in the U.S.A. over the
period 1979-1995.

The data considered here are the M1 money stock (M1) and the Producer Price Index for all commodities ($PPI_{ACO}$). The M1 represents the money supply and $PPI_{ACO}$ the inflation from the point of view of producers. The M1 index is provided by the Board of Governors of the Federal Reserve System while the $PPI_{ACO}$ index is provided by the US Department of Labor. The data are taken from 04/1979 to 12/1995 with a monthly frequency and are available on the web site of the Federal Reserve Bank of St. Louis (Series ID : M1 and PPIACO). The length of the series is $T=200$.

\begin{figure}
    \centering
       \includegraphics[scale=0.65]{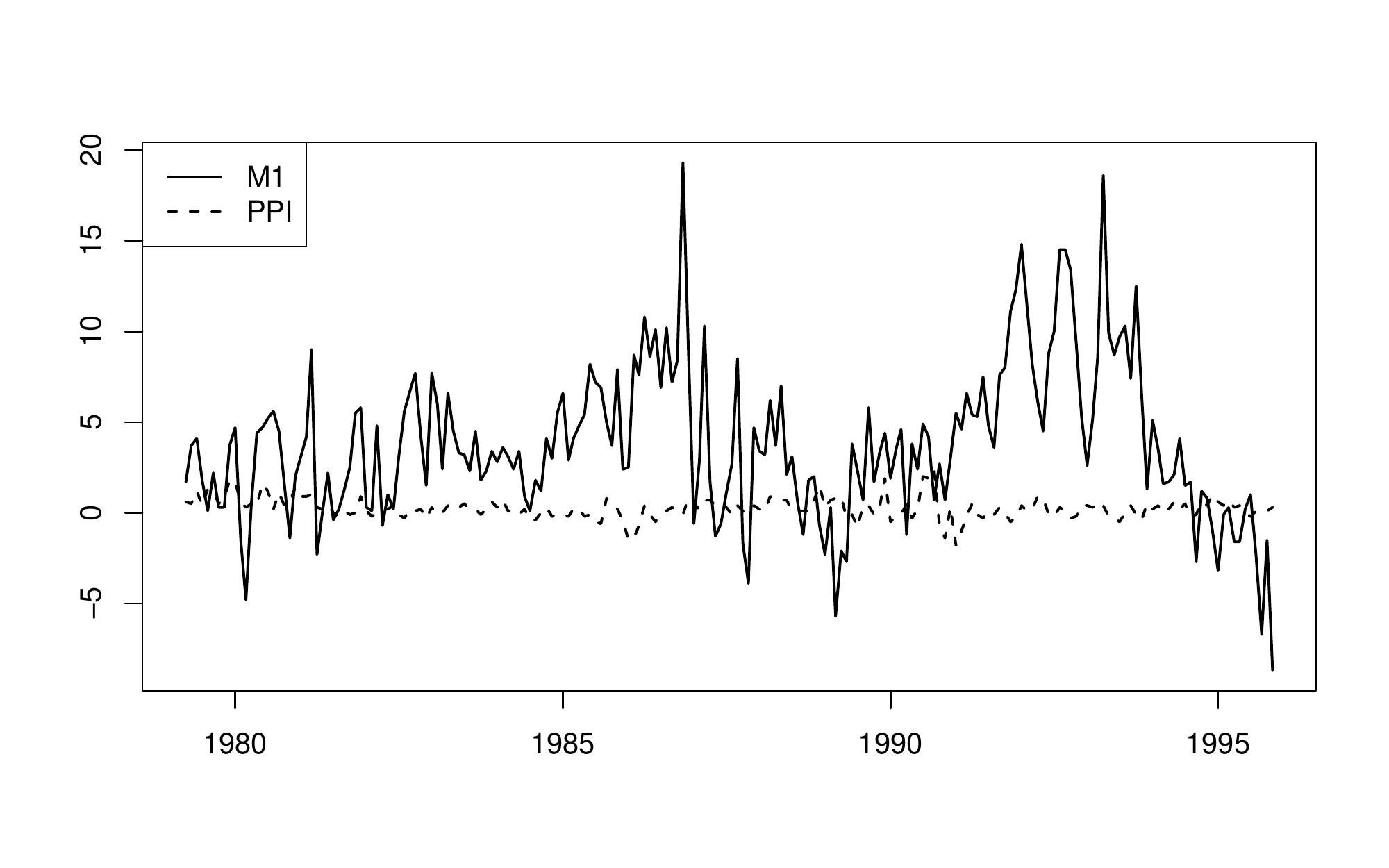}
       \caption{Evolution of the variations $\Delta M1$ and $\Delta PPI_{ACO}$.}
	\label{EvolDatM1}
\end{figure}

The first differences of the data are considered in the sequel. From Figure \ref{EvolDatM1} it appears that the obtained series have non constant variance.
We adjusted a VAR(1) model to the first differences of the series. The autoregressive order is chosen by using portmanteau tests adapted to our non standard framework where the variance structure $\Sigma(r)$ is time-varying (see Patilea and Ra\"{\i}ssi (2011) for details). The outcomes in Table \ref{BoxPierce1} suggest that the model is well fitted. The estimation of the model by the OLS method is given in Table \ref{Param1}. The residuals
of this estimation are next recovered to implement the tests studied in this paper.
Note that we used 399 bootstrap iterations for the $W_b$ test.

\begin{figure}
    \centering
       \includegraphics[scale=0.65]{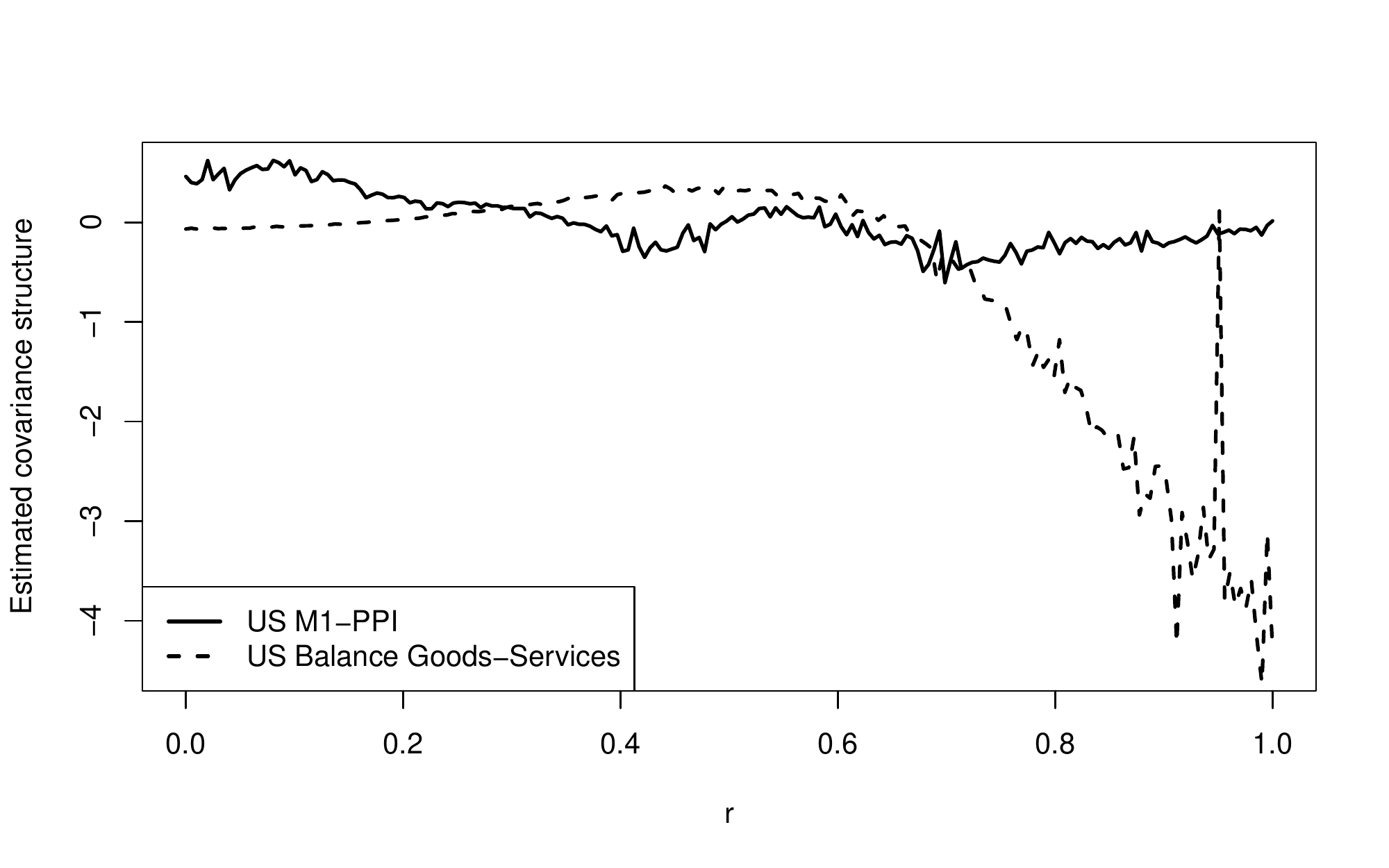}
       \caption{The Nadaraya-Watson kernel estimation of the covariance structure $\Sigma^{12}(r)$ for the two data sets. The estimator is defined as in Patilea and Ra\"{\i}ssi (2012) which showed that such estimator is consistent under {\bf A1} unless at the break points.}
	\label{Covar}
\end{figure}

From Table \ref{Pval1} we see that the $p$-value of the $W_{b}$ test is quite different from those of the $W_{st}$ and $W_{w}$ tests. For instance the null hypothesis of no instantaneous causality is rejected by the $W_{b}$ test for a significance level of $10\%$ while it is accepted by the other tests.
These observations can be explained by the covariance structure of the innovations. Indeed the
nonparametric estimation of this covariance structure plotted in Figure \ref{Covar}
shows that $\Sigma^{12}(r)$ seems not null over the considered period while its seems that $\int_0^1\Sigma^{12}(r)\approx0$.

\begin{center}
\begin{table}[h!]
\centering
{\small
\begin{tabular}{|c|c|c|c|c|c|c|c|c|c|c|c|c|c|c|}
\cline{2-4}
\multicolumn{1}{c|}{} & \multicolumn{3}{c|}{Number of lags}\\
\cline{2-4}
\multicolumn{1}{c|}{} & 3 & 6 & 12\\
\hline
$BP_{OLS}$ & 0.4384 [1.1198]& 0.8169 [1.9071]& 0.8165 [3.2870]\\
\hline
\end{tabular}
}
\caption{The $p$-values of the Box-Pierce test adapted to our non standard framework. The corresponding statistics are displayed into brackets. The $BP_{OLS}$ corresponds to the portmanteau test based on the OLS proxies of the $u_t$'s.
}
\label{BoxPierce1}
\end{table}
\end{center}

\begin{center}
\begin{table}[h!]
\centering
{\small
\begin{tabular}{|c c|c|c|c|c|c|c|c|c|c|c|c|c|}
\cline{1-2}
\multicolumn{2}{|c|}{$\widehat{A}_{01}$}\\
\hline
 0.643 [0.064] & -1.124 [0.360] \\
 -0.009 [0.007] & 0.439 [0.102]\\
\hline
\end{tabular}
}
\caption{The OLS estimators of the matrix $A_{01}$ (see equation (\ref{VAR})) for the adjusted VAR(1) model. Standard deviations of the parameters are displayed into brackets.}
\label{Param1}
\end{table}
\end{center}

\begin{center}
\begin{table}[h!]
\centering
{\small
\begin{tabular}{|c|c|c|c|c|c|c|c|c|c|c|c|c|c|c|}
\cline{2-4}
\multicolumn{1}{c|}{} & $W_{st}$ & $W_{w}$ & $W_{b}$\\
\hline
$p$-values & 0.268 [1.225] & 0.201 [1.632] & 0.058 [10.54]  \\
\hline
\end{tabular}
}
\caption{The $p$-values of the $W_{st}$, $W_{w}$ and $W_{b}$ tests. The corresponding test statistics are displayed into brackets.}
\label{Pval1}
\end{table}
\end{center}

\subsubsection{Merchandise trade balance and balance on services in the U.S.A.}

The merchandise trade balance and the balance on services can be seen as indicators of the economic health of a country. The U.S. merchandise trade balance is the account which redraws the value of the exported goods and the value of the imported goods. The U.S. balance on services is similarly the account which redraws the value of the exported services and the value of the imported services. Here, we search to quantify if it exists an instantaneous causality relation between these two macroeconomic indicators. The data are provided by the Bureau Analysis of the U.S. Department of Commerce and go from the 01/1960 to the 01/2011 with quarterly frequency. The length of the series is $T=204$. They are available on the web site of the Federal Reserve Bank of St. Louis (Series ID : BOPBM and BOPBSV).

\begin{figure}
    \centering
       \includegraphics[scale=0.65]{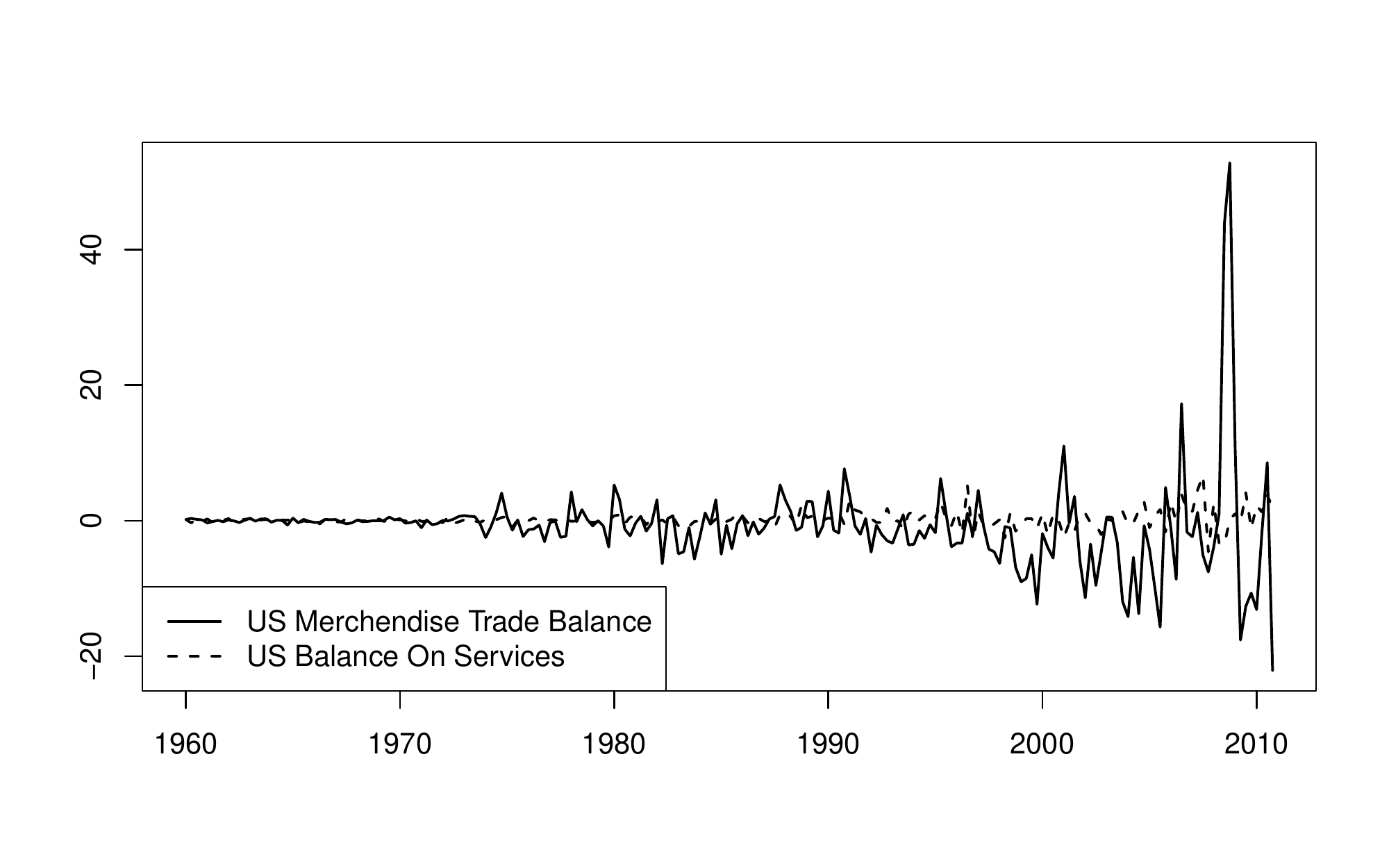}
       \caption{Evolution of the first differences of the U.S. merchandise trade balance and the U.S. balance on services in billion of U.S. dollars.}
	\label{EvolDatBal}
\end{figure}

Similarly to the first data set, we consider the first differences of the data (see Figure \ref{EvolDatBal}).
A VAR(2) model is adjusted to the data (estimation results not reported here).
The adequacy of the model is again checked using portmanteau tests which are
valid in our framework. The portmanteau test suggests to choose a VAR(2) model.
Indeed the $p$-value of the $BP_{OLS}$ test is 0.65[5.29] and
for 5 autocorrelations in the portmanteau statistics (the portmanteau statistic is given into brackets). The $p$-values of the three tests are next computed from the residuals as for the first data set. The outcomes displayed in Table \ref{Pval2} show that the considered tests have quite different results. In view of the non constant variance of the studied series (see Figure \ref{Covar}), the result corresponding to the $W_b$ test is more reliable.

\begin{center}
\begin{table}[h!]
\centering
{\small
\begin{tabular}{|c|c|c|c|c|c|c|c|c|c|c|c|c|c|c|}
\cline{2-4}
\multicolumn{1}{c|}{} & $W_{st}$ & $W_{w}$ & $W_{b}$\\
\hline
$p$-value & 0.0498 [3.848] & 0.341 [0.907] & 0.441 [190.142]  \\
\hline
\end{tabular}
}
\caption{The $p$-values of the $W_{st}$, $W_{w}$ and $W_{b}$ tests. The corresponding test statistics are displayed into brackets.}
\label{Pval2}
\end{table}
\end{center}

\section{Conclusion}
\label{conclusion}

In this paper we studied the problem of testing instantaneous causality in the important case where the unconditional variance is time-varying. The properties of the Wald tests based on the assumption of constant unconditional variance are investigated in our non standard framework. It emerges that this kind of tests may have no power in this non standard framework. As a consequence we proposed a new bootstrap test for testing the instantaneous causality hypothesis in the important case where the unconditional variance structure is time-varying. In particular we found that the proposed bootstrap test is consistent. We illustrated these theoretical results through a set of numerical experiments. The outcomes of macroeconomic data sets suggest that the Wald test may deliver results which are quite different from the bootstrap test. Although our non-standard framework allows for non constant variance, it assumes that the structural innovations cannot display conditional heteroscedasticity. This case could be the object of interesting further researches.

\section*{Appendix}

The following Lemmas are similar to Lemmas 7.2, 7.3 and 7.4 of Patilea and Ra\"{\i}ssi (2012),
so that the proofs are omitted. Introduce
$v_t=\mbox{vec}(u_{1t}u_{2t}'-\Sigma_{t}^{12})$ with
$u_t=(u_{1t}',u_{2t}')'$ and recall that $\vartheta_t=u_{2t}\otimes
u_{1t}$.

\begin{lem}\label{lem1} Under {\bf A1}
we have
\begin{equation*}
\lim_{T\to\infty}E\left[v_{[Tr]}v_{[Tr]}'\right]=(G_2(r)\otimes G_1(r))
\mathcal{M}(G_2(r)\otimes G_1(r))'-\mbox{vec}(\Sigma^{12}(r))\mbox{vec}(\Sigma^{12}(r))',
\end{equation*}
and
\begin{equation*}
\lim_{T\to\infty}E\left[\vartheta_{[Tr]}\right]=\mbox{vec}(\Sigma^{12}(r))
\end{equation*}
for values $r\in(0,1]$ at which the functions $g_{ij}(r)$ are
continuous.
\end{lem}

\vspace{0.3 cm}

\begin{lem}\label{precedent} Under {\bf A1} we have
\begin{equation}\label{ciel1}
T^{-1}\sum_{t=1}^Tv_{t}v_{t}'\rightarrow\lim_{T\to\infty}T^{-1}\sum_{t=1}^TE(v_{t}v_{t}'),
\end{equation}
and
\begin{equation*}
T^{-1}\sum_{t=1}^T\vartheta_{t}\rightarrow\lim_{T\to\infty}T^{-1}\sum_{t=1}^TE(\vartheta_{t}).
\end{equation*}
\end{lem}

\vspace{0.3 cm}

\begin{lem}\label{lemtcllln} Under {\bf A1}
we have
\begin{eqnarray}\label{ciel2}
T^{-1}\sum_{t=1}^TE(v_{t}v_{t}')&\rightarrow& \int_0^1(G_2(r)\otimes
G_1(r))\mathcal{M}(G_2(r)\otimes G_1(r))'dr\\&-&\int_0^1
\mbox{vec}(\Sigma^{12}(r))\mbox{vec}(\Sigma^{12}(r))'dr\nonumber,
\end{eqnarray}
and
\begin{equation*}
T^{-1}\sum_{t=1}^TE(\vartheta_{t})\rightarrow\int_0^1\Sigma^{12}(r)dr.
\end{equation*}
\end{lem}

\vspace{0.3 cm}

\noindent{\bf Proof of Lemma \ref{lem}}\quad We first give the proof
of (\ref{res1}). Let us define $u_t(\theta)=(u_{1t}'(\theta),
u_{2t}'(\theta))'=X_t-(\tilde{X}_{t-1}\otimes I_d)\theta$ for any
$\theta\in\mathbb{R}^{pd^2}$. From the Mean Value Theorem we have
\begin{eqnarray}\label{mvt}
T^{-1}\sum_{t=1}^T\hat{u}_{2t}\otimes \hat{u}_{1t}&=&T^{-1}\sum_{t=1}^Tu_{2t}\otimes u_{1t}\\&+&
T^{-1}\sum_{t=1}^T\left\{\frac{\partial u_{2t}(\theta)}{\partial\theta'}\otimes
u_{1t}(\theta)+u_{2t}(\theta)\otimes\frac{\partial u_{1t}(\theta)}
{\partial\theta'}\right\}_{\theta=\theta^*}(\hat{\theta}-\theta_0)\nonumber,
\end{eqnarray}
where $\theta^*$ is between $\hat{\theta}$ and $\theta_0$, and
$\partial u_t(\theta)/\partial\theta'=-\tilde{X}_{t-1}'\otimes I_d$
is uncorrelated with $u_t$. Hence we write
$T^{-\frac{1}{2}}\sum_{t=1}^T\hat{v}_t=T^{-\frac{1}{2}}\sum_{t=1}^Tv_t+o_p(1)$
with $v_t=\mbox{vec}(u_{1t}u_{2t}'-\Sigma_{t}^{12})$, since the
estimator $\hat{\theta}$ is such that
$\sqrt{T}(\hat{\theta}-\theta_0)=O_p(1)$. The process $(v_t)$ is a
martingale difference sequence, so that from the Lindeberg central
limit theorem $T^{-\frac{1}{2}}\sum_{t=1}^Tv_t$ is asymptotically
normal with mean zero. The expression of the covariance matrix
$\Omega$ can be obtained as follows, using
$E(\epsilon_t\epsilon_t'|\mathcal{F}_{t-1})=I_d$ and  Lemmas
\ref{lem1} and \ref{lemtcllln}:

\begin{eqnarray}
&&\Omega:
=\lim_{T\to\infty}T^{-1}\mbox{Cov}(\sum_{t=1}^Tv_t,
\sum_{t=1}^Tv_t)
=\lim_{T\to\infty}T^{-1}\sum_{t=1}^TE(v_tv_t')\nonumber\\&=&
\lim_{T\to\infty}T^{-1}\sum_{t=1}^TE\left\{\left(u_{2t}\otimes u_{1t}-\mbox{vec}
(\Sigma_t^{12})\right)\left(u_{2t}\otimes u_{1t}-\mbox{vec}(\Sigma_t^{12})\right)'\right\}\nonumber\\&=&
\lim_{T\to\infty}T^{-1}\sum_{t=1}^T\left[E\left\{(u_{2t}u_{2t}')\otimes
(u_{1t}u_{1t}')\right\}-\mbox{vec}(\Sigma_t^{12})
\mbox{vec}(\Sigma_t^{12})'\right]\label{remark}\\
&=&
\lim_{T\to\infty}T^{-1}\sum_{t=1}^T\left[E\left\{(H_{2t}\epsilon_t\epsilon_t'H_{2t}')
\otimes(H_{1t}\epsilon_t\epsilon_t'H_{1t}')\right\}
-\mbox{vec}(\Sigma_t^{12})
\mbox{vec}(\Sigma_t^{12})'\right]
\nonumber\\&=&\lim_{T\to\infty}T^{-1}\sum_{t=1}^T\left[(H_{2t}\otimes H_{1t})
E\left(\epsilon_t\epsilon_t'\otimes\epsilon_t\epsilon_t'\right)(H_{2t}\otimes H_{1t})'
-\mbox{vec}(\Sigma_t^{12})\mbox{vec}(\Sigma_t^{12})'\right]\nonumber\\&=&
\int_0^1(G_2(r)\otimes G_1(r))\mathcal{M}(G_2(r)\otimes G_1(r))'dr-\int_0^1
\mbox{vec}(\Sigma^{12}(r))\mbox{vec}(\Sigma^{12}(r))'dr,
\nonumber
\end{eqnarray}
where the identity $(F\otimes J)(K\otimes L)=(FK)\otimes(JL)$ is
used for matrices of appropriate dimensions. The proof of
(\ref{res0}) follow directly from Lemmas \ref{lem1},
\ref{precedent}, \ref{lemtcllln} and equation (\ref{mvt}).

For the proof of (\ref{res2}) again note that $\hat{v}_t$ can be replaced by $v_t$
from (\ref{mvt}). We write

\begin{eqnarray*}
v_t&=&u_{2t}\otimes u_{1t}-\mbox{vec}(\Sigma_t^{12})
\\&=&H_{2t}\epsilon_t\otimes H_{1t}\epsilon_t-\mbox{vec}(H_{1t}H_{2t}')
\\&=&(H_{2t}\otimes H_{1t})\{\epsilon_t\otimes\epsilon_t-\mbox{vec}(I_d)\}.
\end{eqnarray*}
Define $v_t^{\epsilon}:=\epsilon_t\otimes\epsilon_t-\mbox{vec}(I_d)$. We have
$E(v_t^{\epsilon})=0$ and $Var(v_t^{\epsilon})=E(\epsilon_t\epsilon_t'\otimes\epsilon_t\epsilon_t')
-\mbox{vec}(I_d)\mbox{vec}(I_d)'=:\tilde{\Omega}$. Therefore from Theorem 3.1 of Hansen (1992) it follows that

$$T^{-\frac{1}{2}}\sum_{t=1}^{[Ts]}v_t\Rightarrow\int_0^s(G_2(r)\otimes G_1(r))dB_{\tilde{\Omega}}(r)$$
for $0\leq s\leq1$.$\quad\square$

\noindent{\bf Proof of (\ref{bootres})}\quad For the sake of simplicity and with no loss of generality (see (\ref{mvt})) let us assume that $X_t=u_t$, so that the error process is observed and there is no autoregressive parameters to estimate. Conditionally on the $u_t$'s, $\delta_s^{(i)}$ is a Gaussian process with independent increments and variance

$$E^*(\delta_s^{(i)}\delta_s^{(i)'})=
T^{-1}\sum_{t=1}^{[Ts]}\vartheta_t\vartheta_t',$$
where $E^*(.)$ is the expectation under the bootstrap probability measure. The result follows if

$$T^{-1}\sum_{t=1}^{[Ts]}\vartheta_t\vartheta_t'\rightarrow
\int_0^s(G_2(r)\otimes
G_1(r))\mathcal{M}(G_2(r)\otimes G_1(r))'dr,$$
uniformly for all $s\in[0,1]$. Since $T^{-1}\sum_{t=1}^{[Ts]}\vartheta_t\vartheta_t'$ is monotonically increasing and the limit function is continuous, it suffices to establish the pointwise convergence following Hansen (2000, proof of Lemma A.10). This holds using similar arguments as for (\ref{ciel1}) and (\ref{ciel2}), see Patilea and Ra\"{\i}ssi (2012) Lemmas 7.3 and 7.4.$\quad\square$

\section*{References}
\begin{description}
\item[]{\sc Amendola, A. and
Francq, C. } (2009) Concepts and tools for nonlinear time series
modelling. Handbook of Computational Econometrics, eds: {\sc D.
Belsley} and {\sc E. Kontoghiorghes}, Wiley.

\item[]{\sc
Andrews, B., Davis, R.A. and Breidt, F.J.} (2006) Maximum likelihood
estimation for all-pass time series models. \textit{Journal of
Multivariate Analysis} 97, 1638-1659.

\item[]{\sc Asafu-Adjaye, J.} (2000) The relashionship between energy consumption, energy prices and economic growth : time series evidence from Asian developing countries. \textit{Energy Economics} 22, 615-625.

\item[]{\sc Ashenfelter, O., and Card, D.} (1982) Time series representations of economic variables and alternative models of the labour market. \textit{Review of Economic Studies} 49, \textit{N.5 Special Issue On Unemployment}, 761-781.

\item[]{\sc Aue, A., H\"{o}rmann, S., Horv\`{a}th, L. and  Reimherr, M.} (2009) Break detection in the covariance structure of multivariate time series models.
\textit{The Annals of Statistics} 37, 4046-4087.

\item[] {\sc Bahadur, R.R. (1960)} Stochastic comparison of tests.
\textit{Annals of Mathematical Statistics} 31, 276-295.

\item[] {\sc Bai, J. (2000)} Vector autoregressive models with structural changes in
regression coefficients and in variance-covariance matrices.
\textit{Annals of Economics and Finance} 1, 303-339.

\item[]{\sc Bauwens, L., Laurent, S., and Rombouts, J.V.K.} (2006) Multivariate
GARCH models: A survey. \textit{Journal of Applied Econometrics} 21,
79-109.

\item[]{\sc Benderly, J., and Zwick, B.} (1985) Inflation, real balances, output, and real stock returns. \textit{The American Economic Review} 75, 1115-1123. 

\item[]{\sc Brovelli, A., Ding, M., Ledberg, A., Chen, Y., Nakamura, R., and Bressler, S.L.} (2004) Beta oscillations in a large-scale sensorimotor cortical network : Directional influences revealed by Granger causality. \textit{Proceedings of the National Academy of Sciences of the United States of America} 101, 9849-9854.

\item[]{\sc Case, K.E., Fair, R.C., and Oster, S.M. } (2011) Principles of macroeconomics, 10th edition. Prentice Hall.

\item[]{\sc Dahlhaus, R.} (1997) Fitting time series models to nonstationary processes. \textit{Annals of
Statistics} 25, 1-37.

\item[]{\sc Dahlhaus, R., and Subba Rao, S.} (2006) Statistical inference for time-varying ARCH
processes. \textit{Annals of Statistics} 34, 1075-1114.

\item[]{\sc Davidson, R., and Flachaire, E.} (2008) The wild bootstrap,
tamed at last. \textit{Journal of Econometrics} 146, 162-169.

\item[]{\sc Deane, G., and Gutmann, M.P.} (2003) Blowin' down the road: investigating bilateral causality between dust storms and population change in the great plains. \textit{Population Research and Policy Review} 22, 297-331.

\item[]{\sc Den Haan, W.J., and Levin, A.} (1997) {\em A practioner's guide to robust
covariance matrix estimation}. Handbook of Statistics 15, Chapter
12, (eds: {\sc G.S. Maddala} and {\sc C.R. Rao}) Amsterdam:
Elsevier, 299-342.

\item[]{\sc Francq, C., and Gautier, A.} (2004) Estimation of time-varying ARMA models
with Markovian changes in regime. \textit{Statistics and Probability Letters} 70, 243-251.

\item[]{\sc Fujita, A., Sato, J.R., Kojima, K., Gomes, R.L., Nagasaki, M., Sogayar, M.C., and Miyano, S.} (2009) Identification and quantification of Granger causality between gene sets. \textit{ArXiv e-prints}, arXiv:0911.1159.

\item[]{\sc Gelper, S., and Croux, C.} (2007) Multivariate out-of-sample tests for Granger causality. \textit{Computational Statistics and Data Analysis} 51, 3319-3329.

\item[]{\sc Gon\c{c}alvez, S., and Kilian, L.} (2004) Bootstrapping autoregressions
with conditional heteroskedasticity of unknown form. \textit{Journal
of Econometrics} 123, 89-120.

\item[]{\sc Gon\c{c}alvez, S., and Kilian, L.} (2007)
Asymptotic and bootstrap inference for AR($\infty$) processes with
conditional heteroskedasticity. \textit{Econometric Reviews} 26,
609-641.

\item[]{\sc Granger, C.W.J.} (1969) Investigating causal relations by
econometric models and cross-spectral methods. \textit{Econometrica}
12, 424-438.

\item[]{\sc Hafner, C. M., and Linton, O.} (2010) Efficient estimation of a multivariate multiplicative
volatility model. \textit{Journal of Econometrics} 159, 55-73.

\item[]{\sc Hamilton, J.D.} (1983) Oil and the Macroeconomy since World War II. \textit{Journal of Political Economy}
91, 228-248.

\item[]{\sc Hamori, S., and Tokihisa, A.} (1997) Testing for a unit root
in the presence of a variance shift. \textit{Economics Letters} 57,
245-253.

\item[]{\sc Hansen, B.E.} (1992) Convergence to stochastic integrals for dependent heterogeneous processes. \textit{Econometric Theory} 8, 489-500.

\item[]{\sc Hansen, B.E.} (1995) Regression with nonstationary volatility.
\textit{Econometrica} 63, 1113-1132.

\item[]{\sc Hansen, B.E.} (2000) Sample splitting and threshold estimation.
\textit{Econometrica} 68, 575-603.

\item[]{\sc Hiemstra, C., and Jones, J.D.} (1994) Testing for linear and nonlinear
Granger causality in the stock price-volume relation.
\textit{Journal of Finance} 14, 1639-1664.

\item[]{\sc Horowitz, J. L., Lobato, I. N., Nankervis, J. C., and
Savin, N. E.} (2006) Bootstrapping the Box-Pierce Q-test: a robust
test of uncorrelatedness. \textit{Journal of Econometrics} 133,
841-862.

\item[]{\sc Horv\`{a}th, L., Kokoszka, P., and Zhang, A.} (2006) Monitoring constancy of variance in conditionally heteroskedastic time series. \textit{Econometric Theory} 22, 373-402.

\item[]{\sc Inoue, A., and Kilian, L.} (2002) Bootstrapping autoregressions with
possible unit roots. \textit{Econometrica} 70, 377-391.

\item[]{\sc Jones, J.D., and Uri, N.} (1986) Money, inflation and causality (another look at the empirical evidence for the USA, 1953–84). \textit{Applied Economics} 19, 619-634.

\item[]{\sc Kim, C.S., and Park, J.Y.} (2010) Cointegrating regressions with time heterogeneity.
\textit{Econometric Reviews} 29, 397-438.

\item[] {\sc Kokoszka, P., and Leipus, R.} (2000) Change-point estimation in ARCH models.
\textit{Bernoulli} 6, 513-539.

\item[]{\sc Lee, B.-S.} (1992) Causal relations among stock returns, interest rates, real activity and inflation. \textit{Journal of Finance}
47, 1591-1603.

\item[]{\sc L\"{u}tkepohl, H.} (2005) \!\textit{New Introduction to Multiple Time Series Analysis}. Springer, Berlin.

\item[] {\sc L\"{u}tkepohl, H., and Kr\"{a}tzig, M.} (2004) \textit{Applied Time Series Econometrics},
Cambridge: Cambridge University Press.

\item[] {\sc Mammen, E.} (1993) Bootstrap and wild bootstrap for
high dimensional linear models. \textit{The Annals of Statistics}
21, 255-285.

\item[] {\sc Mankiw, G., and Taylor M.} (2006) Economics. Cengage Learning EMEA.

\item[]{\sc McConnell, M.M., and Perez-Quiros, G.} (2000) Output fluctuations in the United States:
what has changed since the early 1980's? \textit{ The American Economic Review} 90, 1464-1476.

\item[] {\sc Patilea, V., and Ra\"{i}ssi, H.} (2012) Adaptive estimation of
vector autoregressive models with time-varying variance: application to testing
linear causality in mean. \textit{Journal of Statistical Planning and Inference} 142, 2891-2912

\item[] {\sc Patilea, V., and Ra\"{i}ssi, H.} (2012) Testing second order dynamics for autoregressive processes in presence of time-varying variance. Working paper http://www.eea-esem.com/files/papers/eea-esem/2012/2461/patilea\_raissi.pdf.

\item[] {\sc Patilea, V., and Ra\"{i}ssi, H.} (2011) Portmanteau tests for stable
multivariate autoregressive processes. Working paper   arXiv:1105.3638v2 [stat.ME].

\item[]{\sc Pesaran, H., and Timmerman, A.} (2004) How costly is it to ignore breaks when
forecasting the direction of a time series. \textit{International
Journal of Forecasting} 20, 411-425.

\item[]{\sc Qu, Z., and Perron, P.} (2007) Estimating and testing structural changes
in multivariate regressions. \textit{Econometrica} 75, 459-502

\item[]{\sc Ramey, V.A., and Vine, D.J.} (2006) Declining volatility
in the U.S. automobile industry. \textit{The American Economic
Review} 96, 1876-1889.

\item[]{\sc Renault, E., and Werker, B.J.M.} (2004) Stochastic volatility models with transaction time risk. Working paper, Tilburg University: http://arno.uvt.nl/ show.cgi?fid=10530

\item[]{\sc Reichel, R., Thejll, P., and Lassen, K. } (2001) The cause-and-effect relationship of solar cycle length and the northern hemisphere air surface temperature. \textit{Journal of Geophysical Research} 106, 15635-15641.

\item[]{\sc Robinson, P.M.} (1987) Asymptotically efficient estimation in the presence of
heteroskedasticity of unknown form. \textit{Econometrica} 55, 875-891.

\item[]{\sc Sensier, M., and van Dijk, D.} (2004) Testing for volatility changes in U.S.
macroeconomic time series. \textit{Review of Economics and Statistics} 86, 833-839.

\item[]{\sc Seth, A.K.} (2008) Causal networks in simulated neural systems. \textit{Cognitive Neurodynamics}
2, 49-64.

\item[]{\sc Sims, C.A.} (1972) Money income and causality. \emph{American
Economic Review} 62, 540-552.

\item[]{\sc Sanso, A., Arag\'{o}, V., and Carrion, J.L. } (2004) Testing
for changes in the unconditional variance of financial time series. DEA Working Paper,
Universitat de les Illes Balears.

\item[]{\sc St\u{a}ric\u{a}, C.} (2003) Is GARCH(1,1) as good a model as the Nobel prize accolades would imply?
Working paper.

\item[]{\sc Tsay, R.S.} (1988) Outliers, level shifts, and variance
changes in time series. \textit{Journal of Forecasting} 7, 1-20.

\item[]{\sc Turnovsky, S.J. and Wohar M.E.} (1984) Monetarism and the aggregate economy: some longer-run evidence, \textit{The Review of Economics and Statistics} 66, 619-629.

\item[]{\sc White, H.} (1980) A heteroskedasticity consistent covariance matrix
estimator and a direct test for heteroskedasticity.
\emph{Econometrica} 48, 817-838.

\item[]{\sc Xu, K.L., and Phillips, P.C.B.} (2008)
Adaptive estimation of autoregressive models with time-varying
variances. \textit{Journal of Econometrics} 142, 265-280.
\end{description}

\end{document}